# Characterization, description, and considerations for the use of funding acknowledgement data in Web of Science

Adèle Paul-Hus[1], Nadine Desrochers[1] and Rodrigo Costas[2]

[1] adele.paul-hus@umontreal.ca; nadine.desrochers@umontreal.ca
Université de Montréal, École de bibliothéconomie et des sciences de l'information, C.P. 6128, Succ. Centre-Ville, H3C 3J7 Montreal, Qc, Canada
[2] rcostas@cwts.leidenuniv.nl
Center for Science and Technology Studies (CWTS), Leiden University, P.O. box 905, 2300 AX Leiden, the Netherlands

**Abstract**

Funding acknowledgements found in scientific publications have been used to study the impact of funding on research since the 1970s. However, no broad scale indexation of that paratextual element was done until 2008, when Thomson Reuters' Web of Science started to add funding acknowledgement information to its bibliographic records. As this new information provides a new dimension to bibliometric data that can be systematically exploited, it is important to understand the characteristics of these data and the underlying implications for their use. This paper analyses the presence and distribution of funding acknowledgement data covered in Web of Science. Our results show that prior to 2009 funding acknowledgements coverage is extremely low and therefore not reliable. Since 2008, funding information has been collected mainly for publications indexed in the Science Citation Index Expanded (SCIE); more recently (2015), inclusion of funding texts for publications indexed in the Social Science Citation Index (SSCI) has been implemented. Arts & Humanities Citation Index (AHCI) content is not indexed for funding acknowledgement data. Moreover, English-language publications are the most reliably covered. Finally, not all types of documents are equally covered for funding information indexation and only articles and reviews show consistent coverage. The characterization of the funding acknowledgement information collected by Thomson Reuters can therefore help understand the possibilities offered by the data but also their limitations.

**Keywords**: funding acknowledgements; Web of Science; bibliometrics

## Introduction

The impact of research funding on scientific publications has been the subject of discussions and investigations by the scientometric community for decades. In 1970, Crawford and Biderman conducted an innovative analysis of sponsorship patterns for American social sciences. At the time, most discussions on the impact of research funds on social sciences were based on data of sponsor expenditures – who gave how much to whom. By changing the perspective and using the acknowledged sources of funding from the first footnote of more than 3,400 sociology papers, the authors found an important increase of the share of papers acknowledging financial support and a considerable diversification of funding sources between 1950 and 1968. More than 40 years later, Costas and Yegros-Yegros (2013) corroborated the added value of using funding acknowledgement information to assess the output of a funding organization. In fact, the authors found that more than 50% of all publications funded by the Austrian Science Fund (FWF) were retrieved only through the analysis of the funding acknowledgement information.

Harter and Hooten (1992) investigated the funding status of a scientific publication as a possible indicator of the quality of the research. Indeed, "[o]ne may suppose that the act of funding implies an anticipation by the funding agency that the outcome of the project will be useful and that it will make a contribution to further research; to a solution of a problem; or to demonstration of a method, process, or activity" (Harter & Hooten, 1992, p. 583). The authors examined 391 papers, looking for a statement of funding in the first footnote, in the citations, and in the acknowledgement section. The study found no relationship between the funding status of a paper and the quality or the utility of that paper as measured by citations. Cronin and Shaw (1999) and Zhao (2010, based on 1998 publications) also studied the relationship between the funding status of Information Science (IS) research articles and their impact as measured by citations. In both cases, funding information was obtained looking at the acknowledgement section of papers when financial support was explicitly mentioned. In one case (Cronin & Shaw, 1999), the citedness of a publication appears to be associated with the journal of publication and an author's nationality, but not with funding, while in the other case (Zhao, 2010) the citedness of funded research was substantially higher than that of non-funded research.

In 1993, the Unit for Policy Research in Science and Medicine (PRISM) proposed to develop a bibliographic database of biomedical research papers that included funding information, originally limited to the UK publications. The records included in the database were selected from the Science Citation Index (SCI) and the Social Sciences Citations Index (SSCI) from Web of Science, then-called the Institute for Scientific Information. The publications were examined and the funding acknowledgements indexed by means of a thesaurus developed and maintained by PRISM and forming the Research Output Database (ROD) (Jeschin, Lewison & Anderson, 1995). Using data from the ROD, Lewison and Dawson (1998) analyzed 122,000 UK biomedical papers published in 1988-1992 in order to assess the output of grantholders for evaluation and policy-making purposes. Also using the ROD data, Rangnekar (2005) performed an analysis of the mention of the Multiple Sclerosis Society, as a funding organization, within multiple-sclerosis-related publications in order to study the visibility of the organization, its research orientations, and the research impact of the papers it funded. The presence of funding and its impact on research was also studied for the field of radiology (Mussurakis, 1994), and ophthalmology and related biomedical research (Ellwin, Kroll & Narin, 1996). In both cases, funding information was obtained by the manual examination of acknowledgements.

From August 2008, Web of Science started to add funding acknowledgement data to its records, "Web of Science indexes the Funding Agency and, if available, any grant numbers. We also index the source text from the original article for understanding of the context of the acknowledgement" (Web of Science, 2009). In response to many funding bodies' requirement to formally acknowledge the source of funding that supported the research, Web of Science introduced this new data source to support the following types of analyses:

- "Track the research output and influence for any funding body or a specific grant / research program
- Identify the strategic scope of a funding body
- Identify vested interests
- Identify future funding opportunities
- Support an existing grant application by showing related information and evidence of previous performance" (Web of Science, 2009).

In 2010, Lewison and Markusova published what was, to our knowledge, the first paper to use the funding acknowledgement data indexed by Web of Science in order to study the funding sources (governmental or private sources) of cancer research in Russia. Shortly after, Shapira and Wang (2010; Wang & Shapira, 2011) studied the impact of research funding on the development and trajectory of research in nanotechnology using funding acknowledgement data from Web of Science. Based on 91,500 nanotechnology articles published between 2008 and 2009, they found that most nanotechnology funding is nationally-oriented, but that collaboration across borders participates to the internationalization of the field.

In the following year, Rigby (2011) discussed the limitations of funding acknowledgements related to simple errors and confusion (misspelling of funding bodies or errors in grant numbers), in other words, the lack of standardization of funding information, as well as limitations related to cultural and political factors affecting how researchers acknowledge their funding, and behaviours such as the tendency to "exaggerate the productivity of certain grants" (p. 368). As underlined by Rigby (2011), funding acknowledgement data remain self-declared information and are thus subject to unethical or inconsistent behaviours, either when authors fail to acknowledge funding sources or when, on the contrary, they acknowledge support they did not actually received. The self-declared nature of this data notwithstanding, funding acknowledgement statements are guided by editorial guidelines as well as funding bodies' policies.

Using Web of Science's funding acknowledgement data, Rigby later analysed more than 3,500 papers published in 2009 to investigate the relationship between the count of funding acknowledgements and the impact of papers (Rigby, 2013). Also using Web of Science data, Lewison and Roe (2012) studied research on cancer in India in terms of funding sources and subject most supported by external funding bodies. Markusova, Libkind and Aversa (2012) also used Web of Science's funding acknowledgement data to study the impact of funding of research output in Russia, with large-scale data on funding provided by Web of Science (14,471 publications supported by 1,975 funding agencies and organizations).

The literature shows a recent diversification of Web of Science's funding acknowledgement data usages. For example, Diaz-Faes and Bordons (2014) performed a patterns analysis of the funding acknowledgement paratext in Spanish scientific publications across different disciplines. This analysis was possible due to the availability of the data, with a sample totalling more than 38,000 papers. Lewison and Sullivan (2015) used the funding acknowledgement data for a large-scale analysis of conflict of interest statements. Morillo, Costas and Bordons (2015) showed that a more comprehensive list of publications associated with the Spanish networking research organisation CIBER could be established using the funding acknowledgement data.

It is therefore obvious that since the implementation of the indexation of funding acknowledgement data by Thomson Reuters, scholars have recognized their potential for better understanding research evaluation and scientific communication. Rigby (2011; 2013) highlighted the possibilities offered by the systematic collection of funding data by Web of Science that, among other things, "facilitate[s] a wide range of measures to assess the impact of individual funding bodies" (2011, p. 370). As pointed out by Rigby (2013), Web of Science's funding acknowledgements provide a new dimension to bibliometric data that can be systematically exploited for the purposes of evaluation and understanding scientific practice. However, as stated in Costas and van Leeuwen (2012),

> "the potential use of the F[unding] A[cknowledgement] information is very much dependent on the algorithm developed by Thomson Reuters, which has not yet been explained in detail. Therefore it is not completely clear how and from where Thomson Reuters takes this information, and if this is done systematically in all journals, for all publications, for all disciplines, etc." (p. 1650).

This limitation remains an important hindrance for the full development of funding acknowledgement research; this paper will therefore shed some light on the characteristics of the data and the underlying implications for their use.

More recently, Elsevier's Scopus also started to collect and index funding information. Since July 2013, Scopus records started including funding information, indexed in four fields: FUND-SPONSOR (funding sponsor), FUND-ACR (funding sponsor acronym), FUND-NO (grant number) and FUND-ALL (combining information from the three other funding fields) (van Servellen, 2015). However, contrary to the Web of Science funding acknowledgement data, Scopus does not give access to the full text of acknowledgements, thus restricting the possibilities of acknowledgement data analysis to funding information solely. To the best of our knowledge, no funding acknowledgement study has been performed yet using Scopus data. Our analysis will be limited to Web of Science data, although future research could certainly perform similar analyses on Scopus or any other source covering acknowledgments or funding data.

*Objective*

The main objective of this paper is to analyze and characterize the presence and distribution of funding acknowledgement data covered in Web of Science in order to support future research on the topic.

## Methodological discussion

*Funding acknowledgement data collected by Thomson Reuters Web of Science*

The funding acknowledgement data collected by Thomson Reuters' Web of Science (WoS) is structured in three different fields: 'Funding Text' (FT), 'Funding Agency' (FO) and 'Grant Number' (FG). FT is the full text of acknowledgements, which contains funding information but also all other contributions acknowledged by authors. The FO field contains the names of agencies and organizations that are acknowledged for their funding contribution, and FG contains grant numbers, which are generally associated to the funding agencies and organizations identified in FO. Figure 1 presents a snapshot of an example of the funding acknowledgement data collected in WoS.

**Funding**

| Funding Agency | Grant Number |
|---|---|
| National Natural Science Foundation of China | 11201073 |
| JSPS postdoctoral fellowship | 26.04021 |
| National Science Foundation of Fujian Province | 2015J01003 |
| Program for Nonlinear Analysis and Its Applications | IRTL1206 |

Close funding text

The author would like to thank two referees for their helpful comments and careful corrections on previous versions. Financial support through National Natural Science Foundation of China (No. 11201073), the JSPS postdoctoral fellowship (26.04021), National Science Foundation of Fujian Province (No. 2015J01003), and the Program for Nonlinear Analysis and Its Applications (No. IRTL1206) are gratefully acknowledged.

Figure 1. Snapshot of the information collected by Web of Science for the document
http://dx.doi.org/10.1016/j.jmaa.2015.06.071

The FT corresponds to the acknowledgement text as collected from the original paper. The FO contains the extraction, from the FT, of the agencies or organizations for which funding support is acknowledged, and the FG field identifies the grant numbers mentioned in the FT, which are then associated to the different funders. The content of FO and FG are then derived from the FT. Given the fact that the main source of information for acknowledgements studies is the FT, this paper will focus on the presence of this specific field, rather than the FO and FG fields, which form a byproduct of the processing of the FT.

*Methodological approach*

A large-scale analysis was performed on all publications included in WoS. All documents covered in the Center for Science and Technology Studies (CWTS) in-house Web of Science database, updated up to week 26 of 2015, were initially included. More than 43 million documents were analyzed in the temporal analysis of the presence of any type of FT. Based on these initial results, we focused part of our analysis on the publications from the period 2009-onwards (over 11 million publications). For these publications, additional information on document types as well as languages[1] has been included in the analysis.

These publications were matched with their subject categories as well as with the different Web of Science Core Collection databases in which they are included: Science Citation Index Expanded (SCIE), Social Science Citation Index (SSCI), and Arts & Humanities Citation Index (AHCI). The inclusion of a publication in a database is linked to the attribution of that publication to subject categories in the databases, which is based on the classification of its source (i.e. journal or issue). One publication can be included in anywhere from one to all three databases (SCIE, SSCI and AHCI); moreover, it can be assigned up to six subject categories.

**Results**

Figure 2 presents the distribution of funding text (FT) in WoS, for the period 2005-2015. We chose 2005 in order to include some years prior to 2008, the year that Thomson Reuters started indexing FT data. Even though the data for the current year is not complete yet, the share of publications with FT reaches its highest point in 2015. The figure shows a clear increase in the proportion of publications with FT from 33% in 2009 to 55% in 2015.

The low number of publications containing FT before 2009 can be explained by the fact that Thomson Reuters began the indexation of funding information in August 2008. However, there is still a small share of publications with FT in the years preceding 2008. This presence has been explained by Thomson Reuters as the result of a retrospective indexation of certain publications added to WoS after August, 2008 (M. Edmunds, personal communication, September 4, 2015), which results in a residual coverage of funding acknowledgements before 2009. Given the low presence of FT data before 2009, the rest of the analysis will focus on the years 2009 to 2015, the period for which there is more FT coverage.

---

[1] Multilingual papers were considered as 'English-language papers' if English was among the languages listed for the paper.

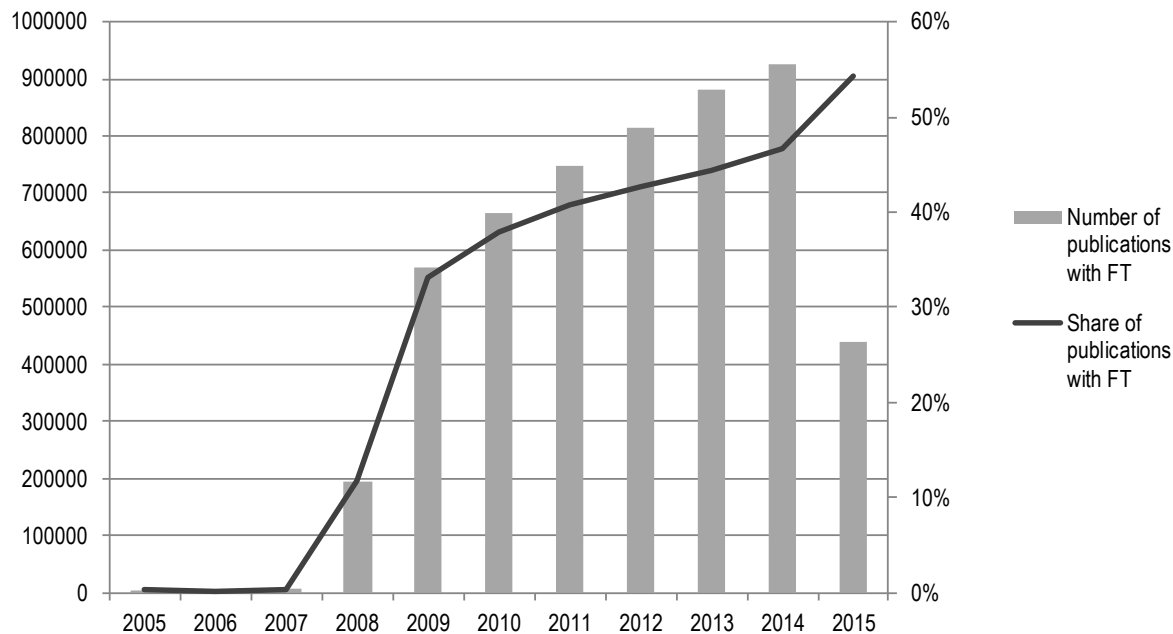

Figure 2. Distribution of FT, 2005-2015

Following the results obtained in Figure 2, we investigated the evolution of funding text coverage by looking at the share of publications containing FT in the different databases, for the period 2009-2015. Figure 3 clearly shows that the biggest share of indexed publications with FT is in the SCIE (alone as well as in combination with either of the two other databases). The share of publications with FT in SCIE steadily increases for the years covered, from almost 40% in 2009 to 60% in 2015. The publications indexed in the SCIE in combination with another database also show increasing shares of FT between 2009 and 2015, from 27% to 39% for the SCIE and SSCI indexed publications, and from 22% to 34% for the SCIE and AHCI indexed publications.

Looking at Figure 3, there is an obvious shift in the coverage of the SSCI publications in 2015; indeed, the SSCI curve clearly shows that prior 2015, funding texts are virtually absent from the publications only indexed in the SSCI. This seems to indicate that the FT indexing of the publications included in the SSCI database began in 2015, which was confirmed by Thomson Reuters (M. Edmunds, personal communication, October 27, 2015). For publications indexed only in the AHCI, the share of publications with FT is stable at 0%. This clearly indicates that there is no coverage of funding acknowledgements included in the sources only indexed in AHCI.

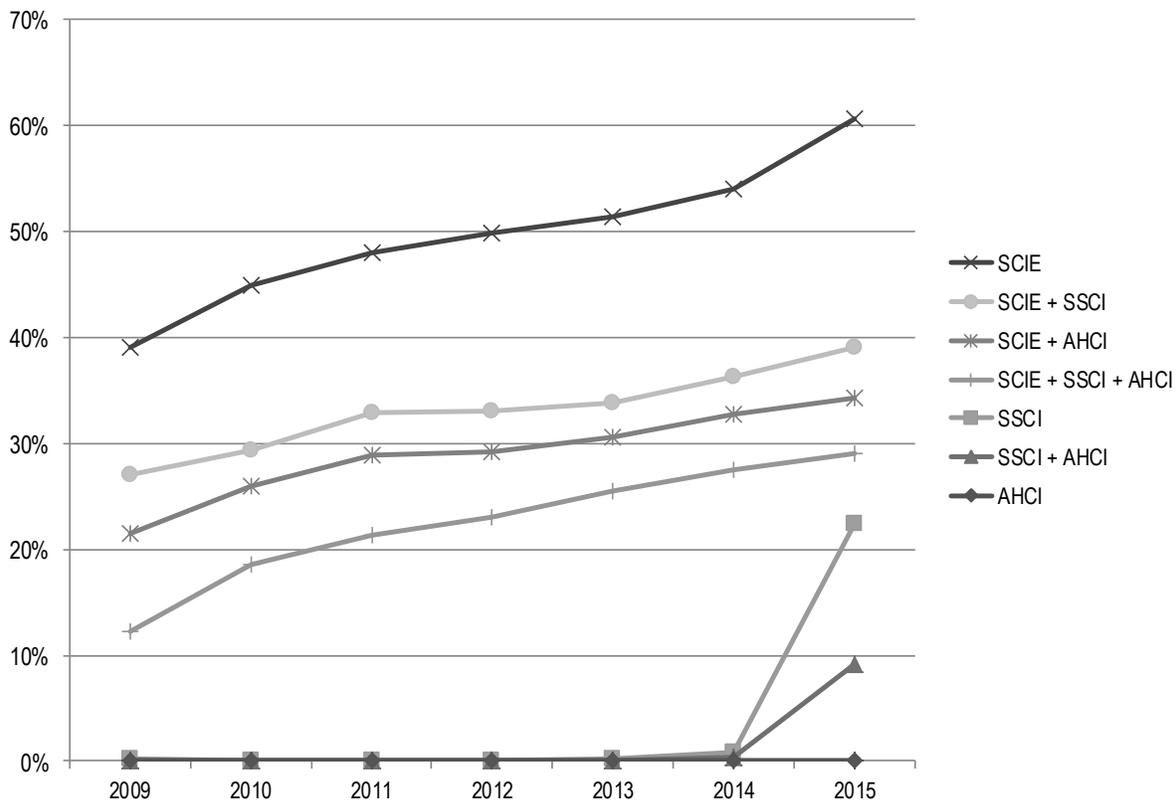

Figure 3. Share of publications with FT by database, 2009-2015

The predominance of the SCIE publications in terms of total number and share of publications with FT seen in Figure 3 is further corroborated when looking at the presence of FT by subject categories, as shown in Figure 4. The map presents the share of publications with FT for all subject categories on a continuum from red to blue where red indicates a high share of FT (up to 80%) and blue an absence of FT. The SCIE covers subject categories from the fields of natural sciences, engineering and medical research (represented mainly on the right-hand side of the map). Figure 3 confirms the dominance of these fields in terms of share of FT, since publications from the Thomson Reuters subject categories of 'Materials science (multidisciplinary)', 'Chemistry (physical and organic)', 'Astronomy & astrophysics', and 'Parasitology' show some of the highest shares of FT. On the other end of the spectrum, Arts and Humanities subject categories (e.g., categories linked to literature, poetry, or history, and found on the left-hand side of the map) show the lowest shares (or complete absence) of FT. The social sciences and medical research publications show shares of FT ranging from 20% to 40%.

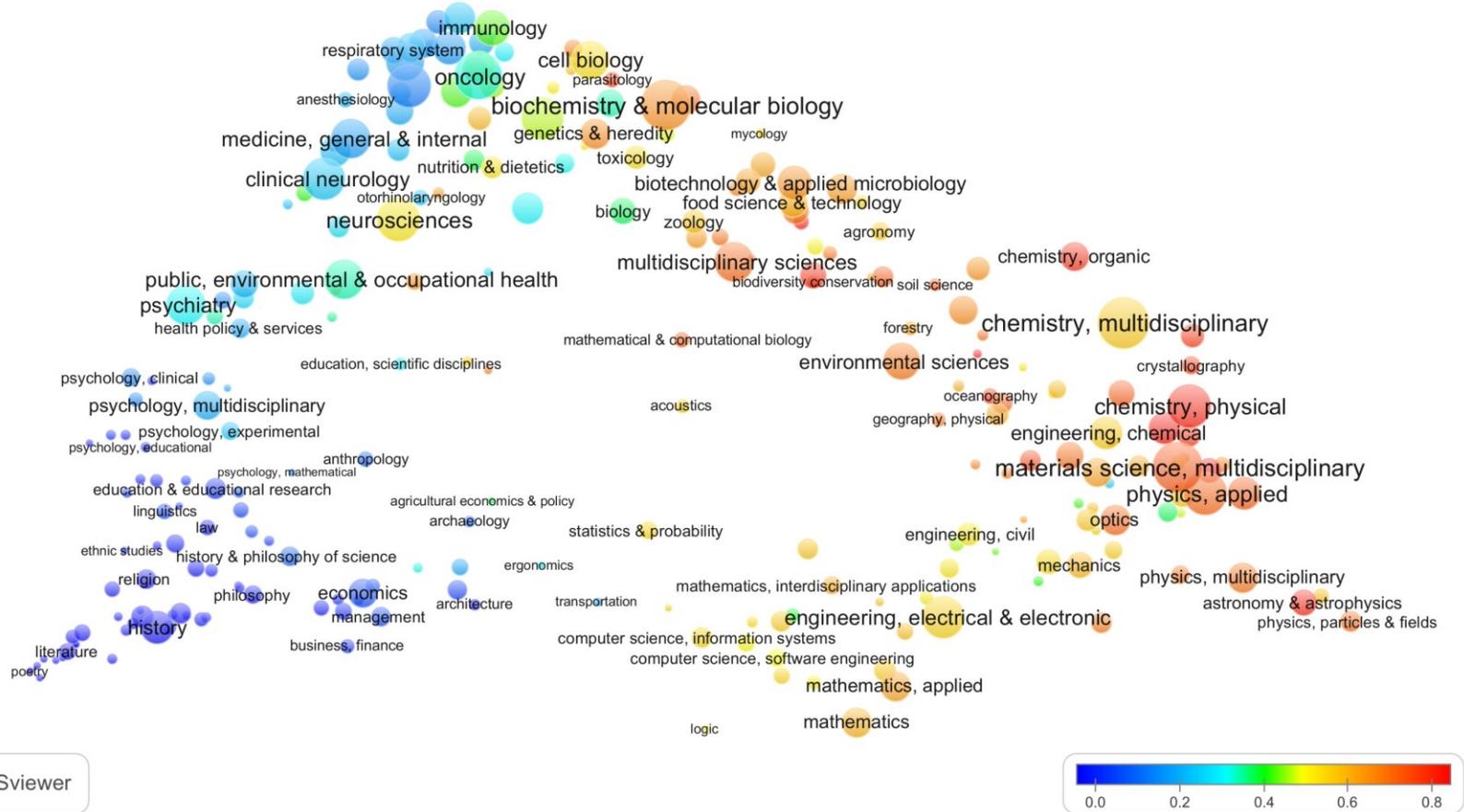

Figure 4. FT presence by subject categories, period 2009-2015

Another consideration in the analysis of funding acknowledgement data in the WoS is document type. Documents labelled as articles and reviews constitute the bulk of publications with FT in WoS databases for the 2009-2015 period. Table 1 shows that these two document types account for more than 99% of all publications with FT; which is more than might be expected given the fact that articles and reviews represent 73% of the total number of publications.

For a more detailed presentation of the presence of FT by database and document type, see Appendix 1; it shows that for all databases, with the exception of the AHCI (which contained essentially no FT for any type of document), there are still cases, though very few, of publications indexed only in the SCIE which contain FT data for document types other than articles and reviews (e.g. Editorial Material or Meeting Abstract).

Table 1. Distribution of FT by document type (2009-2015)

| Document type | Total number of publications | Number of publications with FT | Share of publications with FT |
|---|---|---|---|
| Article | 8341400 | 4815513 | 57.7% |
| Review | 464434 | 223874 | 48.2% |
| Editorial Material | 614024 | 1609 | 0.3% |
| Meeting Abstract | 1511950 | 618 | 0.0% |
| Letter | 280721 | 272 | 0.1% |
| Correction | 84724 | 134 | 0.2% |
| Reprint | 1931 | 46 | 2.4% |
| Book Review | 454166 | 16 | 0.0% |
| News Item | 125344 | 12 | 0.0% |

The distribution of FT by publication language (Table 2) shows the clear dominance of English both in terms of the total number of publications in the WoS (11,432,156 publications for 2009-2015) and the number of publications with FT (5,025,042 publications for 2009-2015). English publications also present the highest share of FT (44%). Despite their lesser presence in terms of total number of publications, more than 35% of publications labelled as published in Chinese contain FT. Publications labelled as published in Spanish, German, or French have a bigger representation than Chinese-language publications when considering the total number of publications in the WoS; however, they each show less than 1% of publications containing FT.

Table 2. Distribution of FT by publication language (2009-2015)

| Language | Total number of publications | Number of publications with FT | Share of publications with FT |
|---|---|---|---|
| English | 11,432,156 | 5,025,042 | 44.0% |
| Chinese | 45,940 | 16,472 | 35.9% |
| Korean | 4197 | 320 | 7.6% |
| Spanish | 85,739 | 61 | 0.1% |
| Portuguese | 43,162 | 43 | 0.1% |
| German | 148,350 | 37 | 0.0% |
| French | 123,510 | 29 | 0.0% |

## Discussion

The availability of the funding acknowledgement data by Thomson Reuters has opened new possibilities of research on acknowledgements (Desrochers, Paul-Hus & Larivière, in press) as well as research funding studies (e.g, Lewison & Markusova, 2010; Lewison & Markusova; Shapira & Wang, 2010). However, a broad analysis of the scope and coverage of the funding acknowledgement data indexed by Thomson Reuters, which can inform the boundaries and actual possibilities in the uses of that data, is still missing in the scientific literature.[2] In this paper we provide an extensive analysis of the presence and distribution of funding text (FT) across years, databases, subject categories, document types, and languages to better understand the characteristics, distribution, and possibilities of these data. In order to properly understand and discuss the results obtained through our analysis, we communicated with Thomson Reuters in order to discuss and confirm our results.

*Years, databases and documents types*

The analysis of the distribution of FT by years of publication shows a very clear pattern. Full coverage in terms of FT data for articles and reviews covered in the SCIE started in 2009; from this date forward, there is an increasing presence of FT over time. For the years before 2009, the proportion of FT-bearing publications is limited and incomplete. For 2008, the low numbers can be explained by the fact that the coverage began in August of that year.

The current Thomson Reuters bibliographic policy states that inclusion of funding acknowledgements should be limited to SCIE and SSCI (Thomson Reuters Bibliographic

---

[2] After the submission of this paper and during the review process, we became aware of another, as-of-yet unpublished study made available through the ArXiv.org repository and addressing the same topic but with important variations in the analysis: cf. Tang, et al. (2016).

Policy Funding Acknowledgements[3], 2015). Publications indexed only in the AHCI are therefore not covered for FT processing. Furthermore, the processing of SSCI publications for funding texts was only introduced in 2015 (M. Edmunds, personal communication, August 27, 2015; see also Figure 3).

The analysis by document type shows that articles and reviews are most likely to include FT. In discussions with Thomson Reuters, the following coverage policy for document types was confirmed: only FT from articles and reviews are indexed for the SCIE while all document types are indexed for the SSCI (M. Edmunds, personal communication, August 27, 2015; Thomson Reuters Bibliographic Policy Funding Acknowledgements, 2015). The fact that the indexing of FT for the SSCI only began recently (2015) further explains the small number of document types other than articles and reviews showing FT.

*Subject categories*

The analysis of the subject categories further shows the dominance of the SCIE-covered publications in terms of FT processing. Moreover, according to Thomson Reuters bibliographic policy (Thomson Reuters Bibliographic Policy Funding Acknowledgements, 2015), publications indexed only in the AHCI are not covered for FT. This explains the lack of FT data for many of the humanities fields.

*Languages*

The analysis of the FT distribution by publication language shows the clear predominance of English and the extremely low number and proportion of FT for publications in Spanish, Portuguese, German or French (that are otherwise amongst the most important languages of publications in WoS, after English). It is important to note that, according to Thomson Reuters' Bibliographic Policy, funding acknowledgements are processed only if they are published in English, regardless of the language of the body of the publication (Thomson Reuters Bibliographic Policy Funding Acknowledgements, 2015). Therefore, the fact that publications in Spanish, Portuguese, German, and French present an extremely low share of indexed FT cannot be taken to mean that they do not contain FT, but rather indicate that they do not contain funding acknowledgement text in English; this, then, contrasts with the high share of publications in Chinese (35.9% of Chinese-language publications) which include English-language funding acknowledgement text.

---

[3] This is an internal guideline policy document. Thomson Reuters indicated to us that they intend to update their website to include these guidelines (Mathilda Edmunds, Director of Content Management, personal communication). In the meantime, interested researchers are invited to contact Thomson Reuters directly for further information.

*Data collection issues*

As the names of the different funding fields suggest, funding acknowledgements are processed and indexed in WoS only if they include mentions of funding. This guideline notwithstanding, the 'Funding Text' field includes the full text of acknowledgements and not just funding related information. Conversely, however, acknowledgements that do not include any mention of funding are not to be indexed (Thomson Reuters Bibliographic Policy Funding Acknowledgements, 2015, p. 6). Therefore, according to the Thomson Reuters bibliographic policy, the set of acknowledgement paratext found in WoS is restricted to funding mention-bearing publications. This is obviously of capital importance for the broader study of acknowledgements, since any analysis of non-funding types of contributions (e.g., data collection, technical assistance, critical reading) based of WoS funding acknowledgements data is biased by the fact that researchers using these data only have access to acknowledgements of non-funding contributions included in acknowledgement texts where funding is also acknowledged.

*General recommendations for researchers and practitioners*

The results presented in this paper have important implications for the study and analysis of acknowledgements, as well as the funding of scientific research. Below are potential considerations, as well as avenues for further research:

1. The first year with a substantial coverage for the SCIE-covered articles and reviews is 2009. The coverage of previous years is extremely low, suggesting a fragmentary coverage, and is therefore not reliable. Furthermore, coverage for this database has increased through the years. Complementary analyses could be conducted to better understand the underlying reasons for this increase and its ramifications for research use of the data.

2. Funding acknowledgement data are covered for all document types for publications covered in SSCI beginning in 2015; there is no coverage before that year. Therefore, researchers should be careful when using results obtained by combining databases as this could provide a misleading impression of coverage in SSCI prior to 2015.

3. FT from publications only covered in the AHCI is not collected. In other words, acknowledgement studies for the field of arts and humanities based solely on WoS AHCI must be avoided. Again, researchers should be wary of results combining other databases with AHCI.

4. Thomson Reuters' bibliographic policy calls for the indexing of FT if this text is in English—no matter the language of the publication. As a result, caution should be used when analyzing FT data from non-English-language publications. Research should then delineate clearly whether the language of the documents and the language of FT are the same.

5.  Researchers using the WoS FT data for broader analyses of acknowledgements (i.e. not limited to funding-related analyses) should be aware that the indexing of the acknowledgements paratext is based on the inclusion of funding-related content.

In conclusion, understanding the characteristics of the funding data collected by Thomson Reuters is an important step in funding acknowledgement research. Results presented in this paper aim to help researchers and practitioners interested in working with those data to understand their possibilities and limitations. Nevertheless, more research should be conducted to further the findings presented here. This could include quantitative and qualitative analyses of samples of FT from the SCIE over the 2009-2015 period in order to ascertain the full scope and ramifications of the increase in coverage found in the present study. As for the SSCI, more detailed analyses of the presence of FT by document type could be performed. Finally, in-depth studies of the subject category distribution of FT data could shed more light on the possibilities and limitations of using these data for revealing the funding acknowledgement landscapes of various fields and disciplines.

## Acknowledgements

This research was conducted while Rodrigo Costas was a visiting scholar at the Canada Research Chair on the Transformations of Scholarly Communication (CRCTSC-Université de Montréal), supported by the Center for Science and Technology Studies (CWTS-Leiden University) and by funding from the South African DST-NRF Centre of Excellence in Scientometrics and STI Policy (SciSTIP). The study was further supported by the Social Sciences and Humanities Research Council of Canada (SSHRC) and the Fonds de recherche du Québec – Société et culture (FRQSC). The authors also thank Mathilda Edmunds and Deepti Chikkam from Thomson Reuters for discussions about their internal documentation and for their input in the interpretation of the results. Finally, they wish to thank María Bordons from the CCHS-CSIC for her insight on funding acknowledgement data and Philippe Mongeon from EBSI (Université de Montréal) for his input on database analysis.

## Appendix 1. Breakdown of publications with FT by document type and across databases

| Databases | Document type | Total number of publications | Number of publications with FT | Share of publications with FT |
|---|---|---|---|---|
| SCIE | Abstract of Published Item | 2 | 1 | 50.0% |
| SCIE | Art Exhibit Review | 1 | 1 | 100.0% |
| SCIE | Article | 7226943 | 4658979 | 64.5% |
| SCIE | Bibliography | 154 | 2 | 1.3% |
| SCIE | Biographical-Item | 20667 | 1 | 0.0% |
| SCIE | Book Review | 2670 | 1 | 0.0% |
| SCIE | Correction | 76213 | 120 | 0.2% |
| SCIE | Database Review | 17 | 6 | 35.3% |
| SCIE | Editorial Material | 468506 | 1137 | 0.2% |
| SCIE | Letter | 248945 | 147 | 0.1% |
| SCIE | Meeting Abstract | 1360550 | 458 | 0.0% |
| SCIE | Meeting-Abstract | 1 | 1 | 100.0% |
| SCIE | News Item | 112828 | 8 | 0.0% |
| SCIE | Poetry | 1 | 1 | 100.0% |
| SCIE | Reprint | 1577 | 30 | 1.9% |
| SCIE | Review | 429821 | 217399 | 50.6% |
| SCIE + AHCI | Article | 5503 | 2118 | 38.5% |
| SCIE + AHCI | Review | 105 | 21 | 20.0% |
| SCIE + SSCI | Article | 247760 | 135140 | 54.5% |
| SCIE + SSCI | Book Review | 9897 | 5 | 0.1% |
| SCIE + SSCI | Correction | 2677 | 4 | 0.2% |
| SCIE + SSCI | Editorial Material | 34441 | 303 | 0.9% |
| SCIE + SSCI | Letter | 12135 | 120 | 1.0% |
| SCIE + SSCI | Meeting Abstract | 107152 | 96 | 0.1% |
| SCIE + SSCI | Reprint | 116 | 14 | 12.1% |
| SCIE + SSCI | Review | 13354 | 6027 | 45.1% |
| SCIE + SSCI + AHCI | Article | 9192 | 3643 | 39.6% |
| SCIE + SSCI + AHCI | Book Review | 6776 | 1 | 0.0% |
| SCIE + SSCI + AHCI | Editorial Material | 566 | 5 | 0.9% |
| SCIE + SSCI + AHCI | Review | 319 | 91 | 28.5% |
| SSCI | Article | 601008 | 14520 | 2.4% |
| SSCI | Book Review | 137439 | 7 | 0.0% |
| SSCI | Correction | 3878 | 9 | 0.2% |
| SSCI | Editorial Material | 59555 | 151 | 0.3% |
| SSCI | Letter | 7934 | 5 | 0.1% |
| SSCI | Meeting Abstract | 44126 | 64 | 0.2% |
| SSCI | News Item | 2884 | 4 | 0.1% |
| SSCI | Reprint | 122 | 2 | 1.6% |
| SSCI | Review | 15906 | 303 | 1.9% |
| SSCI + AHCI | Article | 47712 | 709 | 1.5% |
| SSCI + AHCI | Book Review | 46807 | 2 | 0.0% |
| SSCI + AHCI | Correction | 272 | 1 | 0.4% |
| SSCI + AHCI | Editorial Material | 5390 | 13 | 0.2% |
| SSCI + AHCI | Review | 1211 | 4 | 0.3% |
| AHCI | Article | 202579 | 14 | 0.0% |